\documentclass[pre,amsmath,amssymb,reprint]{revtex4-1}
\usepackage{graphicx}
\usepackage{bm}
\usepackage{hyperref} 
\usepackage[caption=false]{subfig}
\usepackage{comment}
\usepackage{xcolor}

\begin{document}

\title{Electronic parametric instabilities of an ultra-relativistic laser pulse in a plasma}

\author{Ferdinand Gleixner}
\author{Naveen Kumar}\email{naveen.kumar@mpi-hd.mpg.de}
\affiliation{Max-Planck-Institut f\"ur Kernphysik, Saupfercheckweg 1, D-69117 Heidelberg, Germany}
\date{\today}

\begin{abstract}
Electronic parametric instabilities of an ultra-relativistic circularly polarized laser pulse propagating in underdense plasmas are studied by numerically solving the dispersion relation which includes the effect of the radiation reaction force in laser-driven plasma dynamics. Emphasis is placed on studying the different modes in the laser-plasma system and identifying the absolute and convective nature of the unstable modes in a parameter map {spanned by the} normalized laser vector potential and the plasma density. Implications for the ultra-intense laser-plasma experiments are pointed out.
\end{abstract}

\maketitle

\section{Introduction}

Instabilities associated with short-pulse laser propagation in a plasma belong to the so-called parametric instabilities, and they are important for applications in the area of laser driven fusion, laser wakefield acceleration in plasmas, and also for understanding the propagation of light in a medium with refractive index differing from unity~\cite{Kruer:2003yq,Brueckner:1974uq,drake:778,*tripathi:468,mckinstrie:3347,*mckinstrie:2626,Sakharov:1994fk,Sakharov:1997aa,decker:2047,Barr:1999ve,jr.:1440,guerin:2807,*Quesnel:1997aa,*Quesnel:1997ab,Esarey:2009fk}. Since plasma is abound in nature, study of parametric instabilities can also shed light on the physical phenomenon relevant for high-energy astrophysics e.g. cosmic ray acceleration by electromagnetic waves~\cite{Gunn:1969aa,Chen:2002aa,*Chen:2009aa}. 

As the limits for the laser intensities are continued to be pushed further, ultra-high intensity laser systems with $I_{l}\ge 10^{22}\,\text{W/cm}^2$,  are {going to be available} soon~\cite{eli:nn}. These ultra-intense lasers will open new frontiers in the study and application of the laser-plasma interaction~\cite{Marklund:2006aa,*Di-Piazza:2012uq}. In {the} regime of ultra-relativistic laser-plasma interaction, the role of radiation damping in the plasma dynamics becomes important too \cite{Kumar:2013aa}. Concerning the electromagnetic wave propagation in plasmas near astrophysical objects such as nebulae, pulsars etc., {radiation damping in plasmas has to be taken into account} since the electromagnetic waves closer to these objects have large amplitudes~\cite{Gunn:1969aa}. As plasma is a self-organising system, it can support a multitude of modes related to wave propagation. Some of these modes can benefit from the radiation damping of an electromagnetic wave in plasmas and lead to novel effects~\cite{Kumar:2013aa}.

Parametric instabilities of a relativistic laser pulse in a plasma usually consist of four-wave decay interactions; typical examples are Stimulated Raman scattering (SRS), and relativistic modulation instability (RMI) if the perturbation wavevector is restricted to be one dimensional ($1$D), along the laser wavevector direction. For a two-dimensional ($2$D) perturbation wavevector, an additional relativistic filamentation instability (RFI) also becomes important. Among these instabilities, SRS is of significant importance as it is responsible for the generation of hot electrons in fast ignition fusion \cite{Kruer:2003yq,drake:778}, and strong plasma wakefield excitation in laser driven wakefield acceleration~\cite{decker:2047,Esarey:2009fk}.  In a typical four-wave interaction, the incident laser pump decays into two forward moving daughter electromagnetic waves and a plasma wave. The daughter waves have their frequencies upshifted (anti-Stokes wave) and downshifted (Stokes wave) from the laser pump. This {describes} the forward Raman scattering (FRS). In a three-wave interaction,  only the Stokes mode is resonant and its wavevector is either $k_s\approx k_0$ (FRS) or $k_s\approx 2 k_0$ (backward Raman scattering). 

It is instructive to examine the role of the radiation-reaction (RR) force on the plasma dynamics in the ultra-relativistic regime of the laser-plasma interaction, which has been an active area of research in the last few years, especially with regards to the developments of QED-PIC codes~\cite{Gonoskov:2015aa,Arber:2015aa,Derouillat:2018aa}.
In light of this, the effect of the RR force was included in the formalism of electronic parametric instabilities~\cite{Kumar:2013aa}. This formalism is valid in a classical electrodynamics regime where quantum effects {arising} due to photon recoil and spin effects are negligible~\cite{Marklund:2006aa,*Di-Piazza:2012uq}. For this to be valid, the wavelength and magnitude of the external electromagnetic field in the instantaneous rest frame of the electron must satisfy  $\lambda \gg \lambda_C,\, E \ll E_{\text{cr}}$, where $\lambda_C = 3.9 \times 10^{-11}$ cm is the Compton wavelength and $E_{\text{cr}} = 1.3 \times 10^{16}$ V/cm is the critical field of quantum electrodynamics \cite{Di-Piazza:2012uq}. For the laser intensities planned in the ELI project $I_{l}\sim 10^{22-23}\,\text{W/cm}^2$, these two criteria are easily met \cite{eli:nn}. In the classical electrodynamics regime, the Landau-Lifshitz radiation reaction (RR) force  correctly describes the equation of motion for a relativistic charged particle~\cite{Landau:2005fr}. In this paper, we extend the results of Ref.~\cite{Kumar:2013aa} and show that the RR force not only influences the properties of the unstable modes associated with the Raman and modulational instabilities, but also changes the nature of the unstable perturbation in {the} ultra-relativistic regime of laser-plasma interaction. 

The paper is structu{black} as follows: We recall the derivation of the dispersion relation incorporating the leading order term of the Landau-Lifshitz RR force in the plasma dynamics in Sec.\ref{disP}. We then present the results on the temporal analysis of the dispersion relation and identify the various unstable modes and their spectra in Sec.\ref{temP}. This is followed by the parameter maps for number of unstable branches, maximum growth rate and the wavenumber $k$ of the unstable branches in Secs.\ref{maP}, and \ref{gR}. Sec.\ref{spA} shows the spatial properties of the dispersion relation. A parameter map depicting the nature of instabilities is shown in Sec.\ref{conV}. We 
discuss the results in Sec.\ref{conC}.

\section{Dispersion relation}\label{disP}

The details in deriving the dispersion relation can be found in Ref.~\cite{Kumar:2013aa}. We briefly recall here the key points for the sake of continuity;  a circularly polarized (CP) pump laser of strength $a_0=eA_0/m_ec^2$,  $A_0$ is the vector potential of the laser, $e$ is the electronic charge, $m_e$ is the electron mass, and $c$ is the velocity of light in vacuum, propagates in an infinite plasma with uniform plasma electron density $n_0$. The ions are assumed to be at rest. The electrons interacting with the ultra-intense laser pulse experience the radiation damping force that affects the electron dynamics and the laser propagation~\cite{Steiger:1972aa}. The effect of radiation damping on the stationary motion of an electron moving in a circle was analysed in Ref.~\cite{Zeldovich:1975aa} in the limit that the magnetic part of the Lorentz force is compensated by the longitudinal field generated due to the emission of radiation. Incidentally, this is also akin to the Akhiezer-Polovin solution of a circularly polarized laser pulse in an underdense plasma where the magnetic part of the Lorentz force is compensated by the space-charge field of the background plasma~\cite{Akhiezer:1956mz}.  On equating the dominant term of the Landau-Lifshitz radiation friction force with the Lorentz force, approximated as $a_0 m_e c^2$, a threshold vector potential, $a_{\textrm{rad}}$, can be defined as $a_{\textrm{rad}}=\varepsilon_{\textrm{rad}}^{-1/3},\,\varepsilon_{\textrm{rad}}=(4\pi r_e/\lambda_0)$,  where $r_e=e^2/m_e c^2$ is the classical radius of the electron and $ \lambda_0 $ is the wavelength of the laser pulse~\cite{Bulanov:2004aa}. For $a_0 \ll a_{\textrm{rad}}$, the electron motion is described by the usual relativistic motion, $p_{\perp}\approx m_e c a_0$, where $p_{\perp}$ is the transverse momentum. For $a_0 \gg a_{\textrm{rad}}$, the transverse momentum is strongly damped and it scales as $p_{\perp}\approx m_e c  (a_0/\varepsilon_{\textrm{rad}})^{1/4}$~\cite{Bulanov:2004aa}. For a circularly polarized laser pulse of wavelength $\lambda_0=0.8\,\mu$m,  it yields $a_{\textrm{rad}}\approx 280$, corresponding to the laser intensity $I_l \approx 2.5\times 10^{23}$ W/cm$^2$. It is expected to see the onset of radiation reaction effects already at $a_0 < a_{\textrm{rad}}$ in laser-plasma interaction. In Ref.~\cite{Kumar:2013aa}, the effect of radiation reaction force on the equilibrium propagation was analyzed in the limit $a_0\le a_\mathrm{rad}$, approximating $p_{\perp}$ as $p_{\perp}= m_e c a_0(1-i\varepsilon \gamma_0 |a_0|^2)$, where $\varepsilon=2\pi r_e/3\lambda_0$\textcolor{black}{\, $= 7.38 \times 10^{-9}$}, and $\gamma_0=(1+a_0^2/2)^{1/2}$. This equilibrium propagation yields a dispersion relation of the form, $\omega_0^2=k_0^2c^2+{\omega_p^{'2}}\left(1-{i\mu}|A_0|^2\gamma_0/2\right)$, where $\omega_p^{'2}=\omega_p^2/\gamma_0,\, \omega_p = {(4\pi n_0 e^2 /m_e)^{1/2}}$, and $\mu=2e^4\omega_0/3m_e^3c^7$, depicting the damping due to the radiation reaction force~\cite{Kumar:2013aa}; see also \cite{Steiger:1972aa}.  It can be noted that $\mu$ accounts for the radiation reaction induced phase shift and is inversely proportional to the cube power of the electron mass. Thus, the radiation reaction effects are stronger for the plasma electrons than the plasma ions. This justifies ignoring the effect of radiation reaction on ion motion, even for longer laser pulse durations when the ions also start to move.

Apart from the equilibrium propagation of the laser pulse in the plasma, the laser pulse can also suffer stimulated scatterings in a plasma which can be affected by the radiation reaction force~\cite{Kumar:2013aa}. The scattering of the pump laser in a plasma can be written as
\noindent
\begin{multline}
 \bm{A}=\frac{1}{2}\Big[\bm{A}_0 e^{i\psi_{0}}+\bm{\delta A}_{+}e^{i\psi_{+}} +\bm{\delta A}_{-}^{*}e^{-i\psi^{*}_{-}} \Big]+c.c,
\end{multline}
where $\bm{A}_0=\bm{\sigma}A_{0},\bm{\sigma}=(\hat{\bm x}+i\hat{\bm y})/\sqrt{2},\,\text{and}\,\psi_0=k_0 z-\omega_0 t$, $\bm{\delta A}_{+}= \bm{\sigma} \delta A_{+}$, $\bm{\delta A}_{-}^{*}=\bm{\sigma} \delta A_{-}^{*}$, $\bm{\delta A}_{+}$ and $\bm{\delta A}_{-}^{*}$ represent the anti-Stokes and the Stokes waves (Raman sidebands) respectively, $\psi_{+}=(k_z+k_0)z-(\omega+\omega_0)t,\,\psi_{-}^{*}=(k_z-k_0)z-(\omega^{*}-\omega_0)t,\,\omega_0$ and $k_0$ are the carrier frequency and wavevector of the pump laser while $\omega (\omega^{*})$ and $k_z (k_z)$ are the frequency and the wavevector of the anti-Stokes (Stokes) wave, respectively. Beating of these Stokes and anti-Stokes waves with the pump laser field leads to excitation of plasma oscillations, $\delta n/ n_0$,
\noindent
\begin{equation}
\delta \tilde{n}=\frac{e^2 k_z^2}{2m_e^2\gamma_0^2c^2 D_e}\left({A}_{0}^{*}{\delta A}_{+}+{A}_{0}{\delta A}_{-}\right),
\label{plasmaosc}
\end{equation}
where $ D_e=\omega^2-\omega_p^{'2}, \delta n/n_0=\delta \tilde{n}e^{i\psi}e^{i\bm{k}_{\perp}.\bm{x}_{\perp}}/2+c.c$ and $\psi\equiv \psi_{+}-\psi_0 \equiv \psi_{-}^{*}+\psi_0=k_z z -\omega t$. 

The damping of the pump laser field can be incorporated by defining a frequency shift  of the form $\omega_0=\omega_{0r}-i\delta\omega_0,\,\delta\omega_0 \ll \omega_{0r}$ with the frequency shift $\delta\omega_0$ being $\delta\omega_0={\omega_{p}^{'2}\varepsilon\gamma_0 a_0^2}/{2\omega_{0r}} $. Finally solving the wave equation for the scattered vector potentials yields the dispersion relation
\noindent
\begin{equation}
 \left(\frac{R_{+}}{D_{+}}+\frac{R_{-}}{D_{-}}\right)=1,
 \label{disp_rel}
\end{equation} 
where
\begin{multline}
D_{\pm}=(\omega \pm \omega_0)^2-\omega_{p}^{'2}(1\mp i\varepsilon a_0^2\gamma_0)-(k_z\pm k_0)^2c^2,\nonumber \\
R_{\pm}=\frac{\omega_p^2 a_0^2}{4\gamma_0^3}\Bigg[\frac{k_z^2 c^2}{D_e}\left(1\mp i\varepsilon a_0^2\gamma_0 + \frac{2 i \varepsilon a_0^2\gamma_0}{k_z c}  \frac{\omega\omega_0}{\omega\pm\omega_0}\right)\nonumber \\ -(1\pm 4 i \varepsilon \gamma_0^3)\Bigg].
\end{multline}
\noindent
This form of dispersion relation differs from the dispersion relation derived before due to the presence of the radiation reaction term. Without the radiation reaction term $\varepsilon=0$, it assumes the same form as derived before~\cite{drake:778,Kruer:2003yq,decker:2047,Gibbon:2005ys}. As mentioned before, the ion motion is not directly affected by radiation reaction force. However, accounting for ion motion leads to the appearance of new modes associated with the ion motion \emph{e.g.} Brillouin scattering~\cite{drake:778}. This coupling of electron and ion motion can make the dispersion relation, Eq.\eqref{disp_rel}, a higher order polynomial in $\omega$. Moreover, the ion motion can weaken the space-charge potential which consequently affect the plasma density oscillations, $\delta n$, leading to the lower growth rate of the Raman instabilities typically seen in the non-linear stage of these instabilities~\cite{decker:2047}.

\section{Temporal analysis of the dispersion relation}\label{temP}

\begin{figure}
\subfloat[\label{sfig:testa}]{%
  \includegraphics[width=0.32\textwidth]{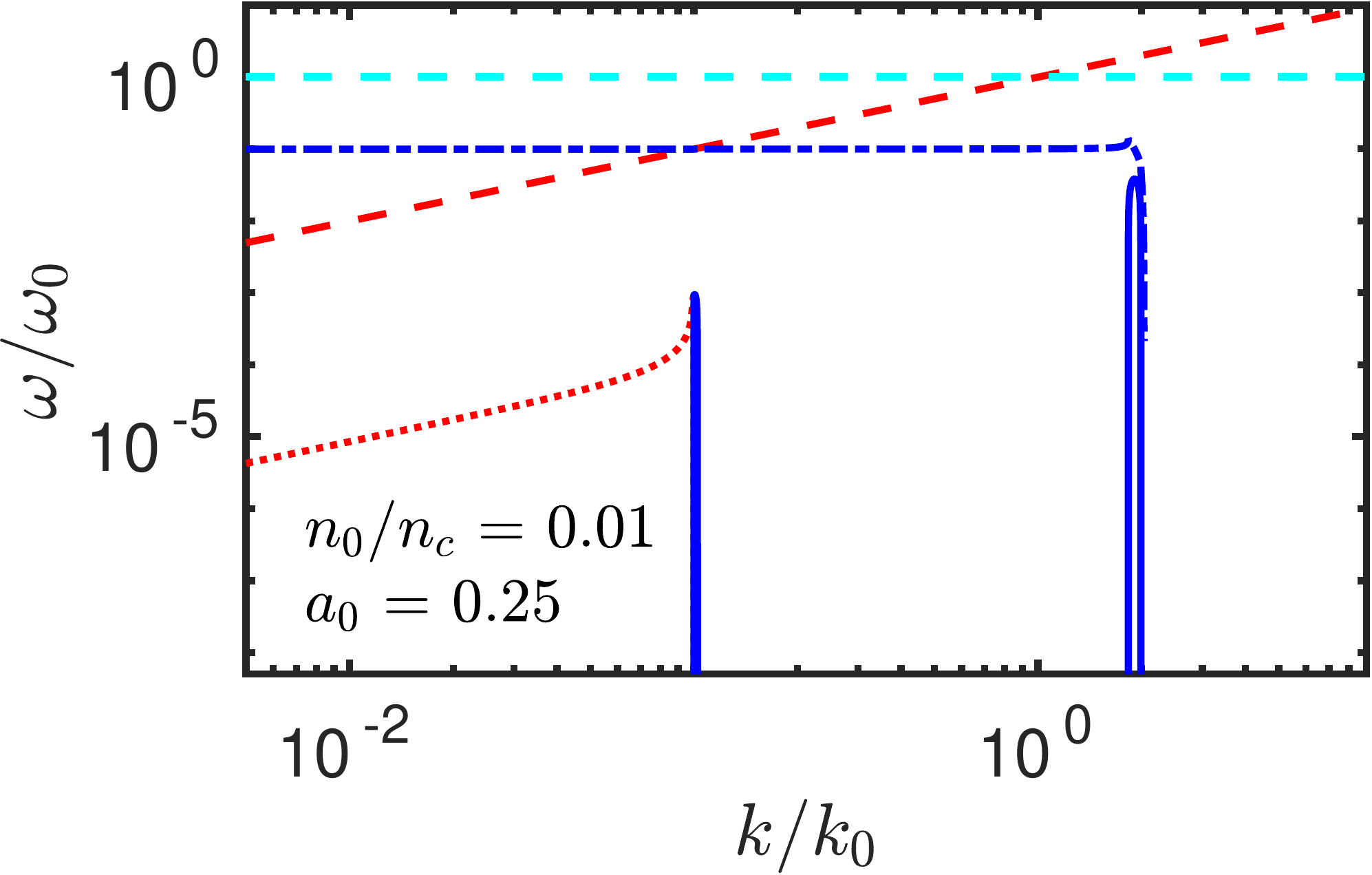}%
}\vfill
\subfloat[\label{sfig:testb}]{%
  \includegraphics[width=0.32\textwidth]{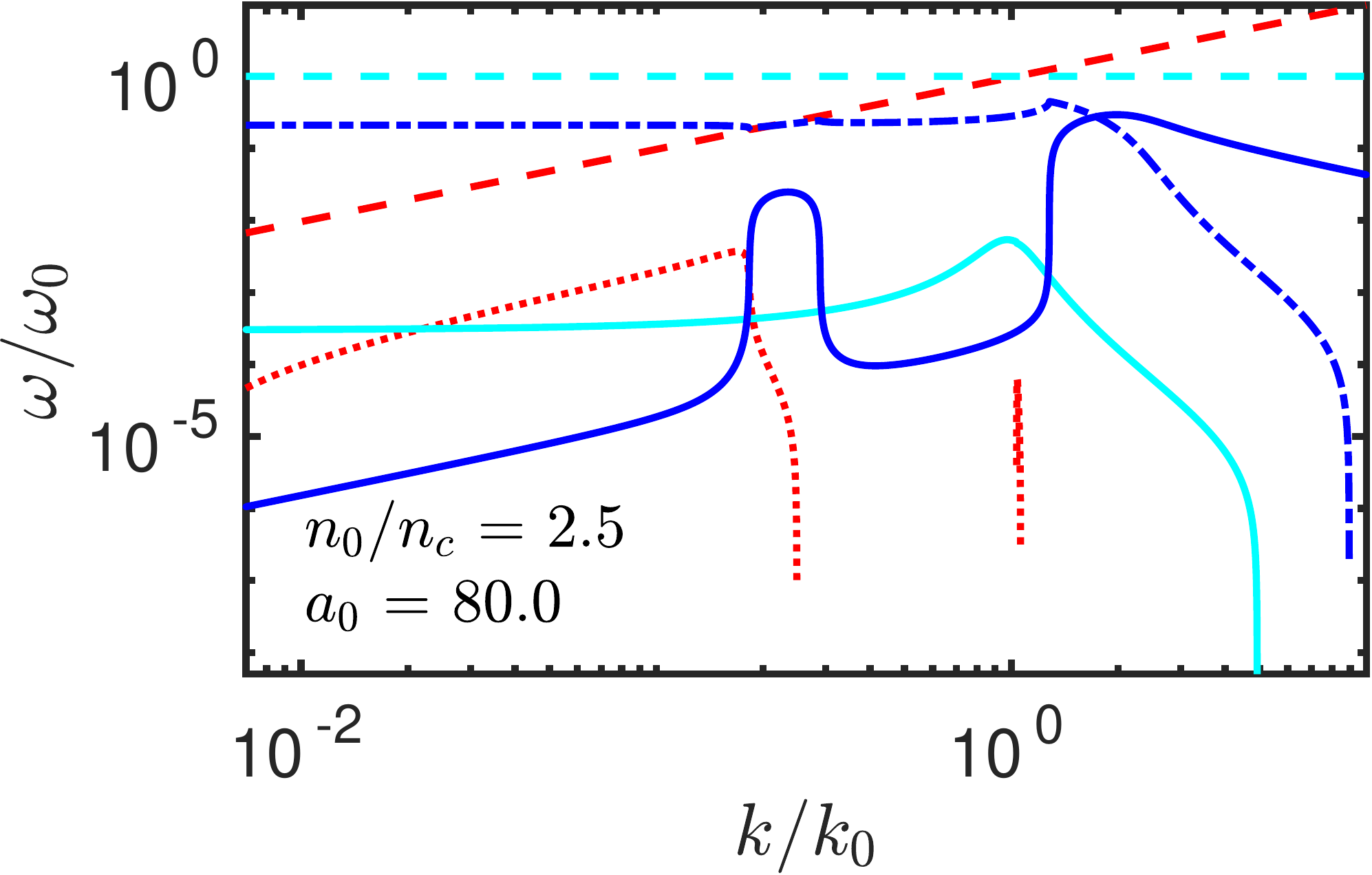}%
}\vfill
\subfloat[\label{sfig:testc}]{%
  \includegraphics[width=0.32\textwidth]{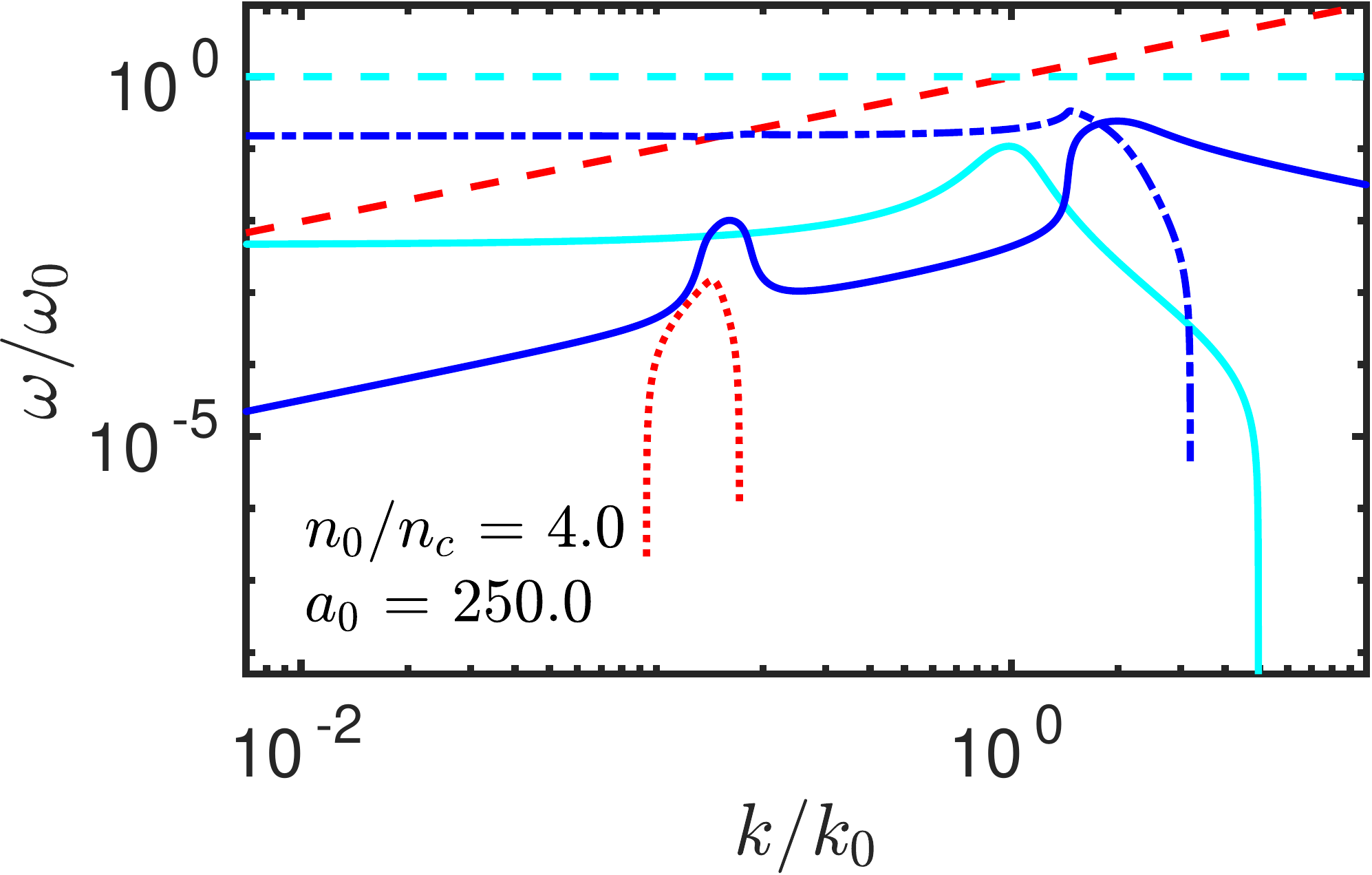}%
}\vfill
\caption{\textcolor{black}{Growth rates of the Raman \textcolor{black}{[solid blue (solid dark grey) line], and modulational [dotted red (dotted light grey) line]} instabilities with the normalized wavevector, $k/k_0$. The \textcolor{black}{dot-dashed blue (dot-dashed dark grey) and the dashed red (dashed dark grey) lines represent the real parts of the corresponding modes, respectively. In panel (a), the dashed cyan line (dashed light grey) represents the laser mode. This laser mode gets unstable and its real part represented by the dashed cyan line (upper dashed light grey) and imaginary part shown by the solid cyan line (lower solid light grey) are depicted in panels (b) and (c). The color legends in the brackets are for the black and white printing of the figures.}} Parameters are (a) $a_0=0.25, n=0.01\, n_c$; (b) $a_0=80, n=2.5\, n_c$; (c) $a_0=250, n=4\, n_c$, where $n_c=m_e\omega_0^2/4\pi e^2$ is the non-relativistic critical density for the laser propagation.}
\label{fig:growth_rates}
\end{figure}
\begin{figure}
\subfloat[\label{sfig:testa}]{%
  \includegraphics[width=0.32\textwidth]{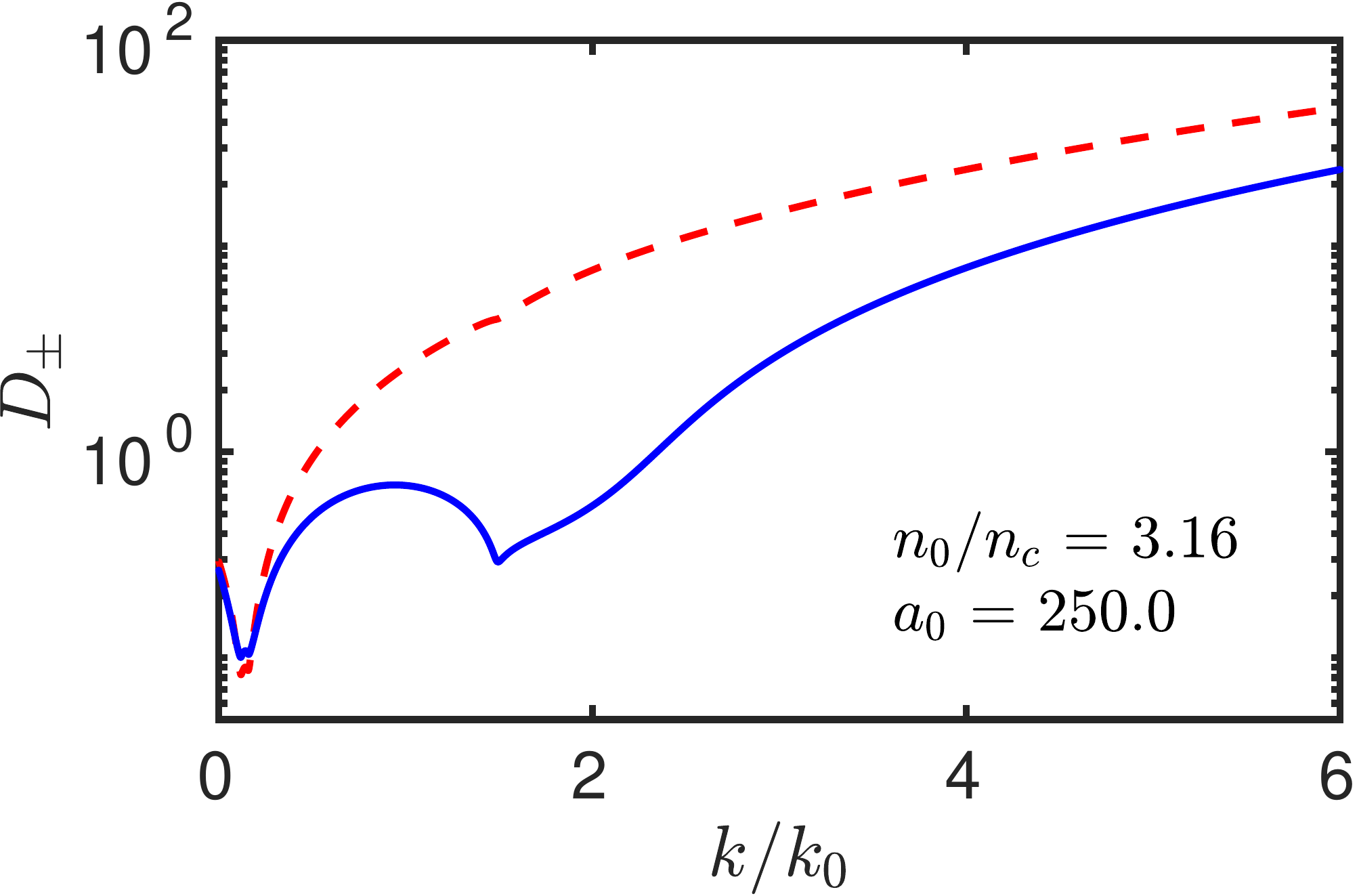}%
}\vfill
\subfloat[\label{sfig:testa}]{%
  \includegraphics[width=0.32\textwidth]{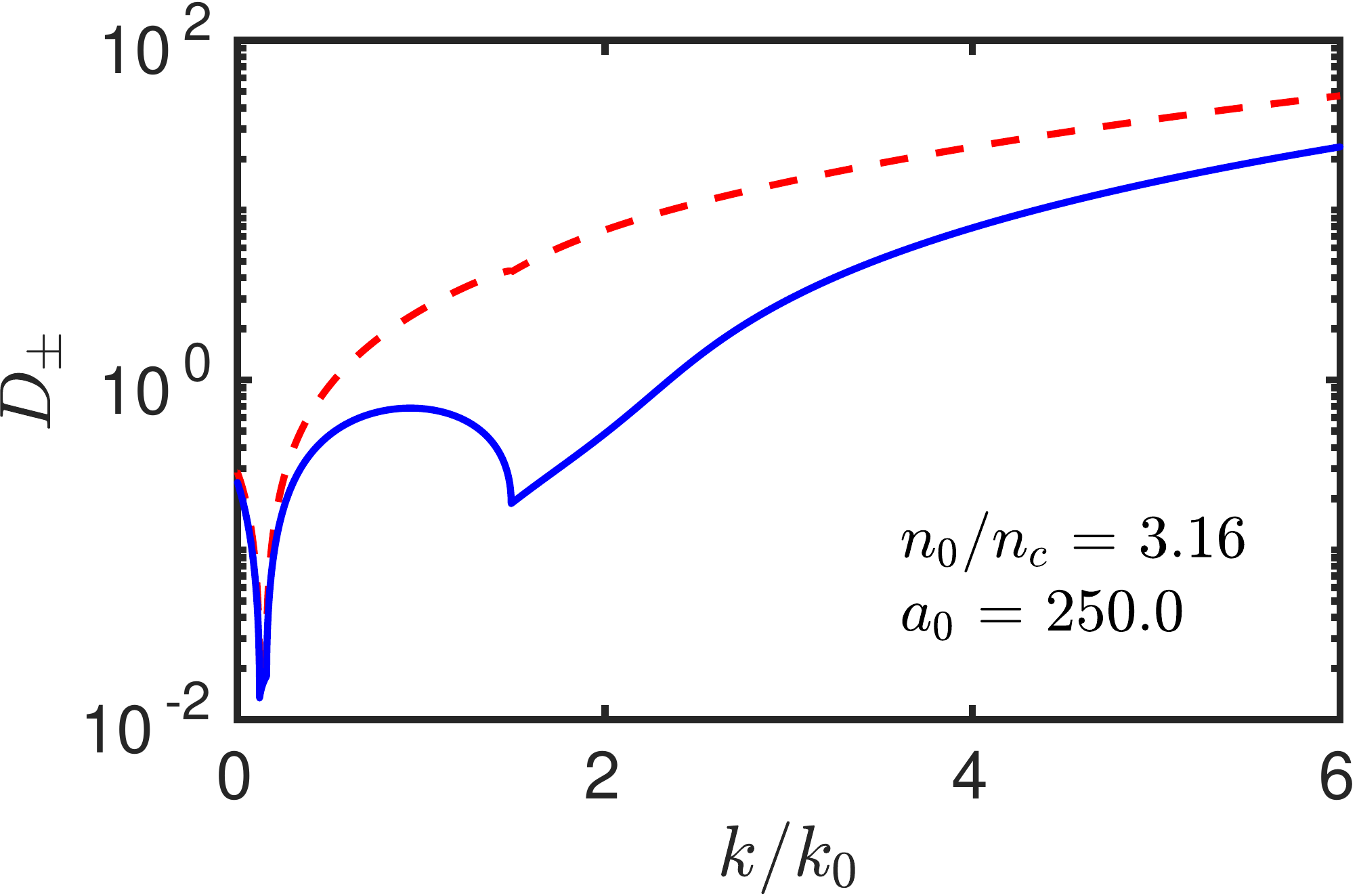}%
}\vfill
\caption{The resonance of the anti-Stokes [$D_+$, \textcolor{black}{dashed red (dashed grey) line}] and Stokes [$D_-$, \textcolor{black}{solid blue (solid dark grey) line}] modes with (a) and without (b) the RR force in a plasma.}
\label{fig:D_pm}
\end{figure}

We solve the dispersion relation, Eq.\eqref{disp_rel}, numerically and follow the evolution of different branches separately. Fig.\ref{fig:growth_rates} shows the two Raman \textcolor{black}{[solid and dot-dashed blue lines, (solid and dot-dashed dark grey)] and the modulational [dotted and dashed red lines (dotted and dashed grey)]} branches. \textcolor{black}{The laser mode is represented by the dashed cyan (dashed light grey) line in panel (a)}.  At lower values of $a_0=0.25$ and density ratio $n/n_c=0.01$, the forward and backward Raman scattering branches are clearly visible with their respective $k$ numbers, in Fig.~\ref{fig:growth_rates}(a). A peak corresponding to the FRS appears at the same location where the real parts of modulation and Raman branches merge \textcolor{black}{[dotted blue (dotted dark grey) and dashed red (dashed dark grey) lines]}. This is expected since at a fixed wavevector $k=k_p$, where $k_p$ is the plasma wavenumber, the modulational interaction assume the form of Raman interaction resulting in the FRS instability. At higher values of $a_0$ and $n/n_c$, these two branches of Raman scattering {merge} into each other. The position of the branches' peaks shift towards each other, suggesting the role of nonlinear frequency shifts becoming important in this process. There is an appearance of another mode \textcolor{black}{[cyan lines (upper dashed and lower solid light grey lines)] in panels(b) and (c) which show} a smaller peak at the laser wavevector $k=k_0$, suggesting that that the laser mode itself gets unstable. While it is known that the inclusion of the radiation reaction force can cause the damping of the laser pulse, the observed growth at the laser wavevector ($k=k_0$) is unexpected. This growth of the laser mode  peaks at the crossover point between the real parts \textcolor{black}{[dotted blue (dotted dark grey) and dashed red (dashed dark grey) lines]} of the modulation branch and laser mode as seen in Fig.\ref{fig:growth_rates}(b), where a spike, related to the modulational branch, is occurring at the same location as the crossover between the real parts of two branches in $k$-space. At higher $a_0=250$, the modulation branch becomes localized in $k$-space and the growth of the laser mode is larger than the modulational branch as seen in Fig.\ref{sfig:testb}. These observations indicate that the growth of the laser mode \textcolor{black}{[solid cyan line (solid lower light grey)]} is most likely a numerical artefact connected with the root-finding algorithm rather than a physical effect. It is the manifestation of the modulation branch being partly merged into the laser mode at high $a_0$. This presumably occurs due to the higher order of the dispersion relation arising in the case of the RR force in Eq.\eqref{disp_rel}. At $a_0=250$, where the RR force effects are important, the FRS and the BRS modes of the Raman branch merges, and consequently only the continuous Raman branch is identified as an unstable mode in a plasma. Although the RR force facilitates bridging the different branches of the Raman instabilities, as apparent from Fig.\ref{fig:growth_rates}, it can also impact the simultaneous resonance of the anti-Stokes and Stokes branches of the Raman modes. On comparing Figs.\ref{fig:D_pm} (a) and (b) it becomes clear that the inclusion of the RR force, [Fig.\ref{fig:D_pm}(a)], can make both  anti-Stokes \textcolor{black}{[solid blue (solid dark grey) line] and Stokes mode [dashed red (dashed grey) line]} of the forward Raman branches simultaneously not fully resonant (defined as $D_{\pm}=0$). This can affect the growth rate enhancement due to the RR force discussed in Ref.~\cite{Kumar:2013aa}.

\subsection{Parameter maps for the number of distinguishable unstable branches}\label{maP}

\begin{figure}
\subfloat[\label{sfig:testa}]{%
  \includegraphics[width=0.4\textwidth,height=0.25\textwidth]{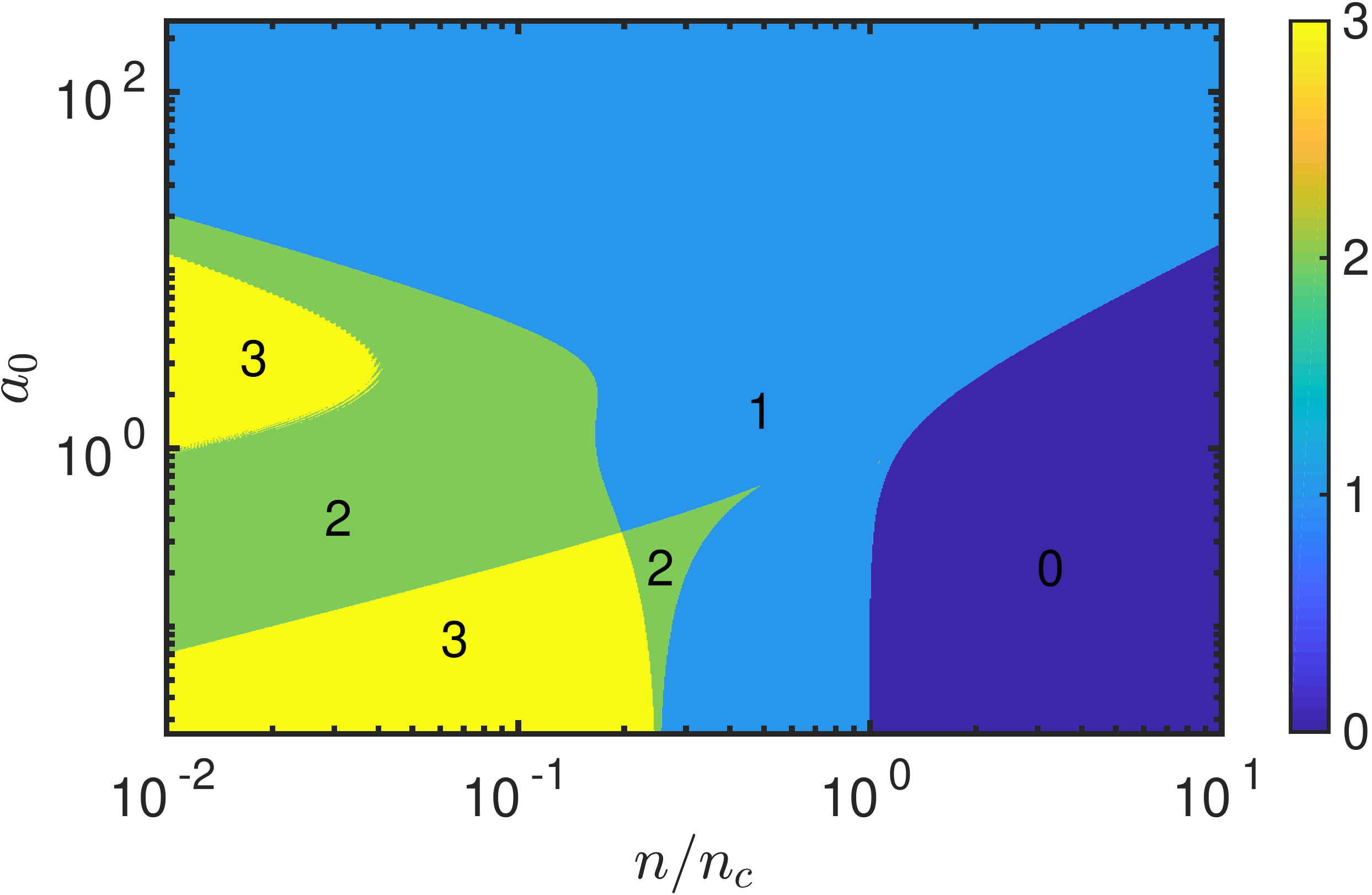}%
}\vfill
\subfloat[\label{sfig:testa}]{%
  \includegraphics[width=0.4\textwidth,height=0.25\textwidth]{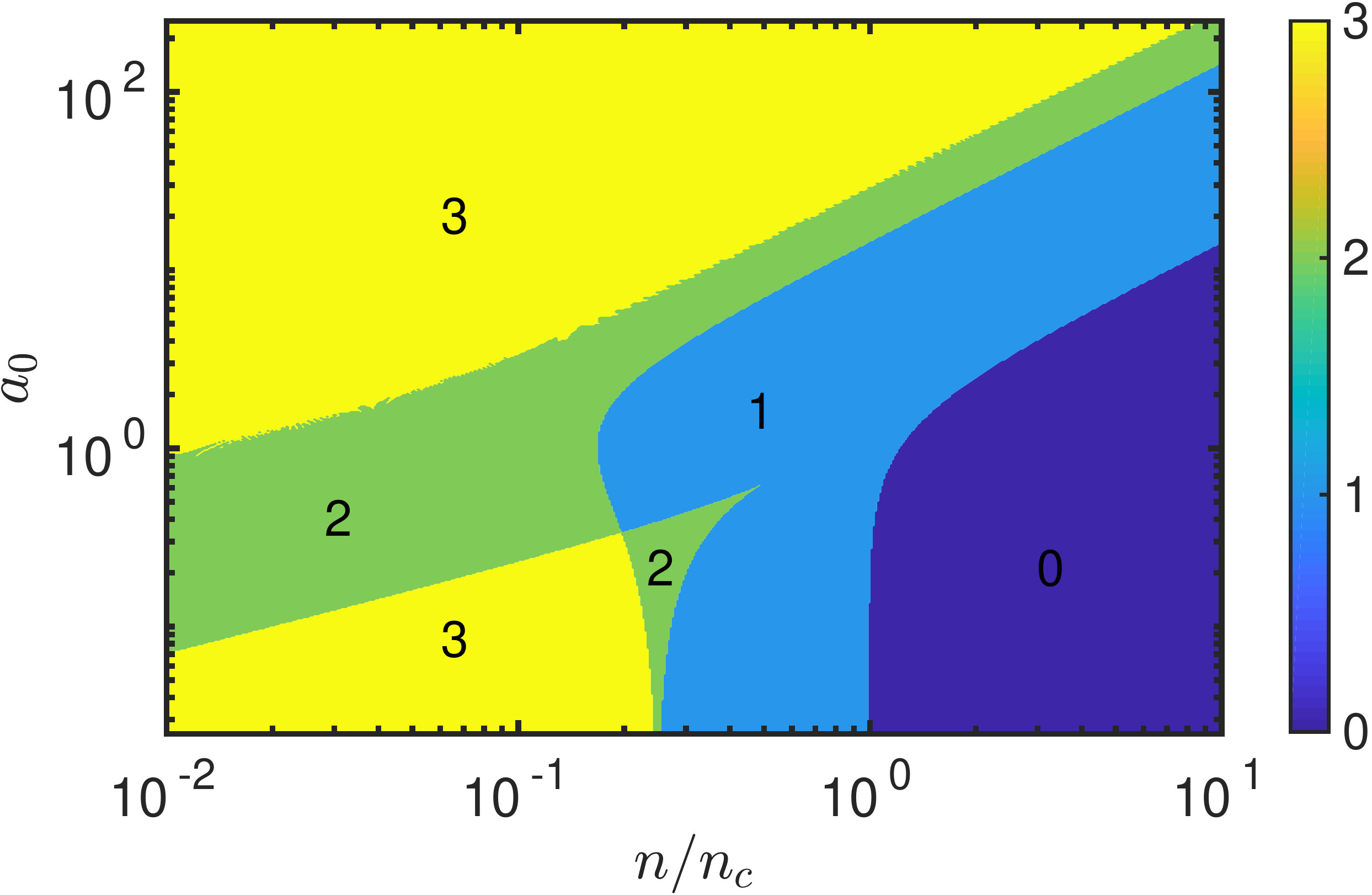}%
}
\caption{Parameter map for the distinguishable unstable branches with (a) and without (b) radiation reaction force. The colorbar denotes the number of distinguishable unstable branches. Dark blue color \textcolor{black}{(dark grey shading)} in the lower right corner denotes the relativistically opaque regions of plasma density.}
\label{fig:branch_parameter}
\end{figure}

The occurrence of various unstable branches is best captured in Fig.\ref{fig:branch_parameter}, which shows the number of distinguishable unstable branches with normalized vector amplitude $a_0$ and the plasma density $n/n_c$. For comparison the unstable branch map, without including the radiation reaction force, is also shown in Fig.\ref{fig:branch_parameter}, reproducing the result of Ref.~\cite{couairon:3434}. On comparing Figs.\ref{fig:branch_parameter}(a) and \ref{fig:branch_parameter}(b), it is clear that the radiation reaction force reduces the number of unstable branches at higher $a_0$. This is consistent with the observation in Fig.\ref{fig:growth_rates}(c).

\subsection{Maximum normalized growth rate and $k$ of the Raman branch of instabilities}\label{gR}

\begin{figure}
\centering
\includegraphics[width=0.4\textwidth,height=0.25\textwidth]{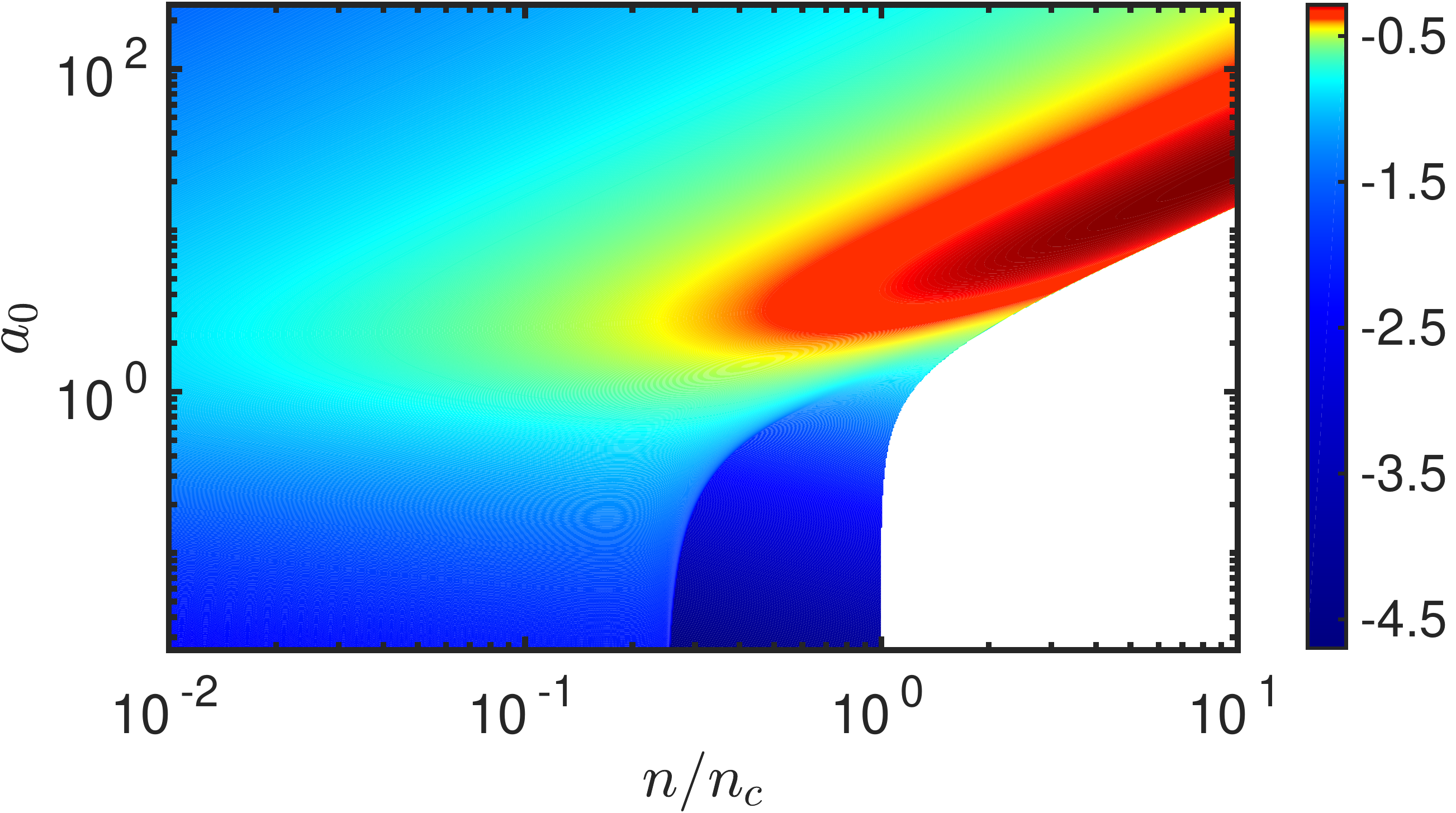}
\caption{Parameter map for the maximum normalized growth rate ($\Gamma/\omega_0$) of the unstable branches with respect to the normalised laser vector potential $a_0$ and the plasma density $n/n_c$. The colorbar is on log$_{10}$ scale. White region in the lower right corner denotes the relativistically opaque regions of plasma density.}
\label{fig:maxgrowth}
\end{figure}

\begin{figure}
\subfloat[\label{sfig:testa}]{%
  \includegraphics[width=0.4\textwidth,height=0.25\textwidth]{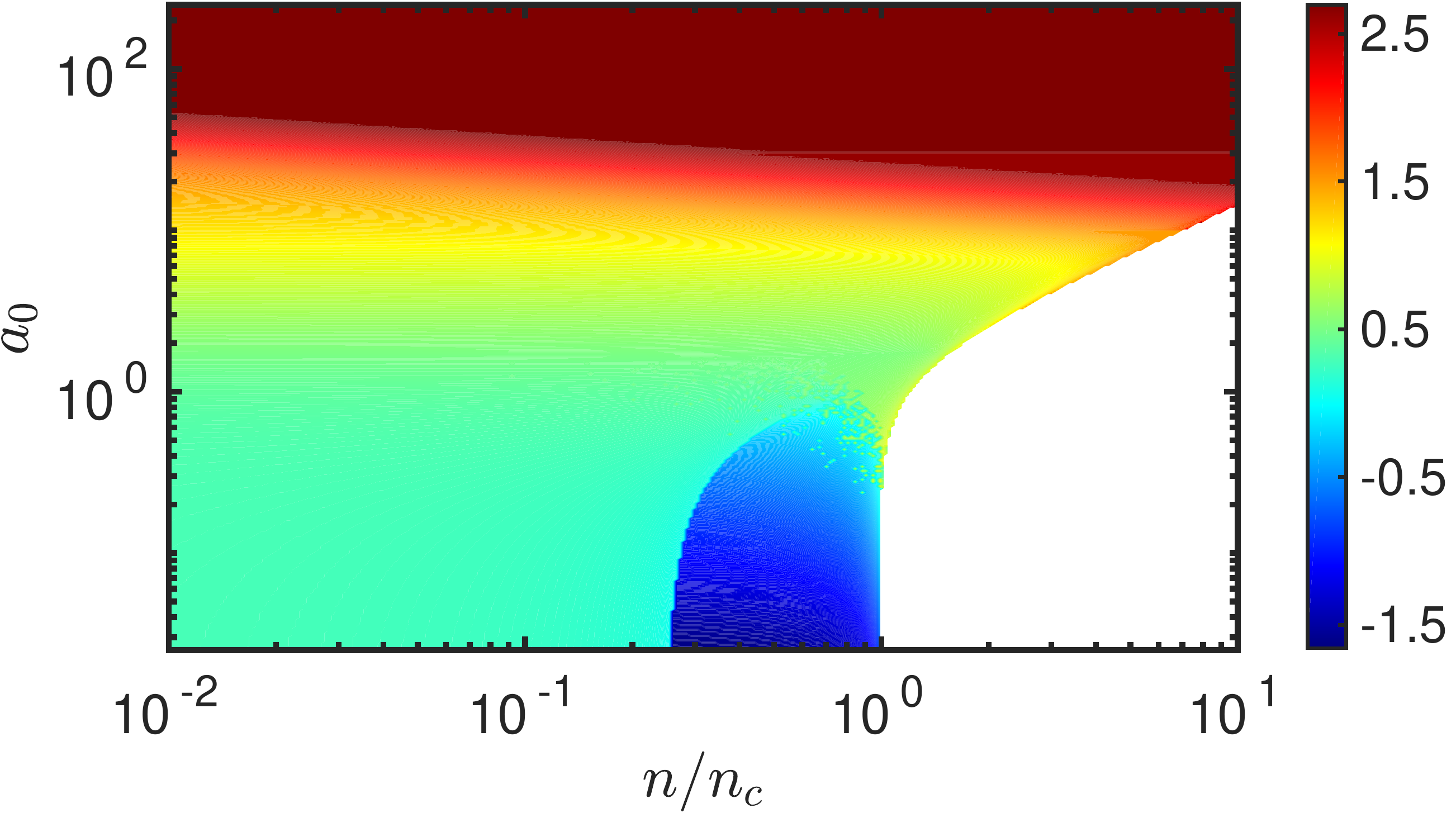}%
}\vfill
\subfloat[\label{sfig:testa}]{%
  \includegraphics[width=0.4\textwidth,height=0.25\textwidth]{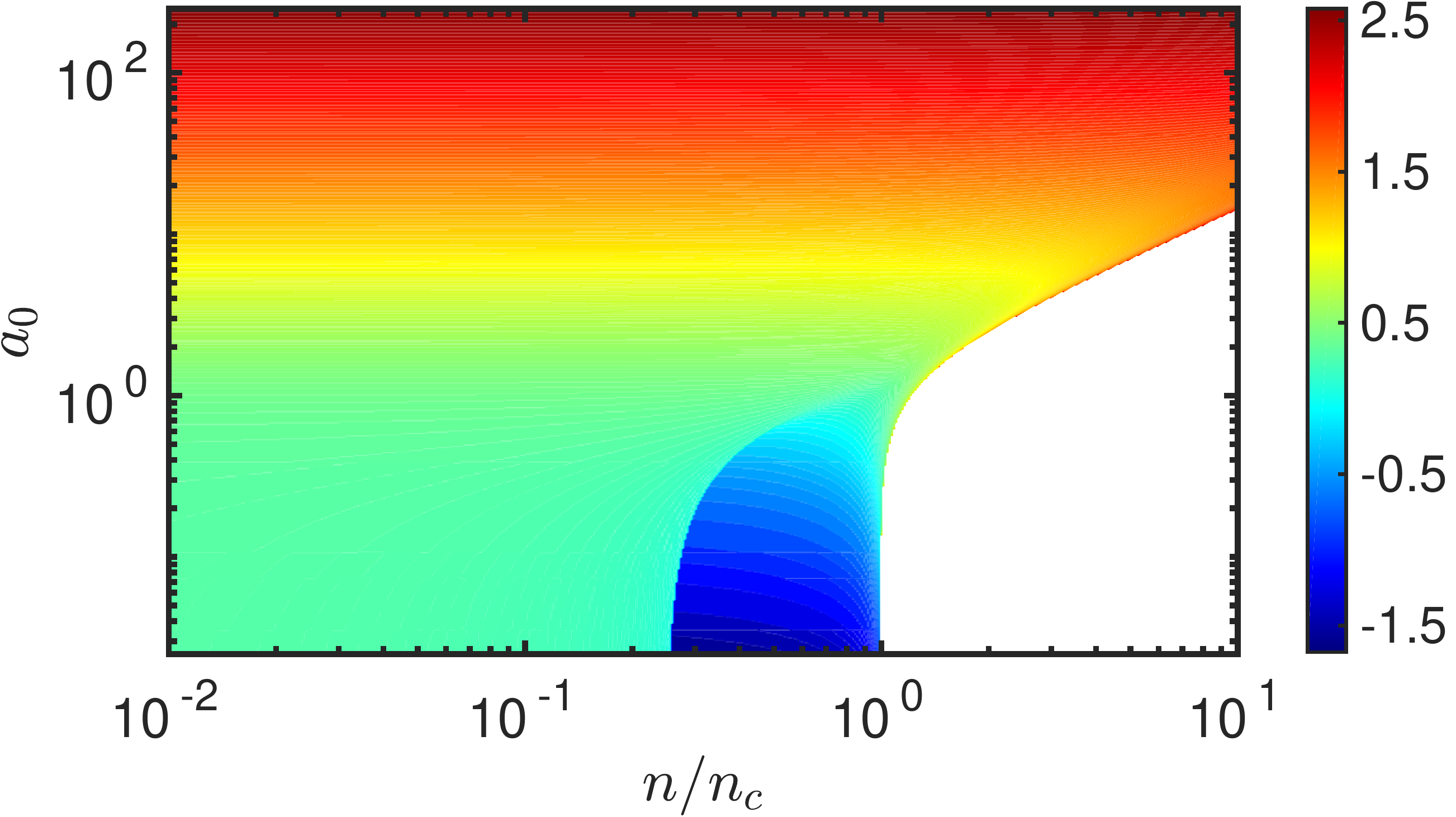}%
}
\caption{Parameter map for the normalized maximum wavevector \textcolor{black}{($k/k_0$)} of any unstable perturbation with (a) without (b) the RR force. The colorbar is on log$_{10}$ scale. White region in lower right corner denotes the relativistically opaque regions of plasma density.}
\label{fig:maxkmap}
\end{figure}

Fig.\ref{fig:maxgrowth} shows the maximum normalized growth rate map of the Raman and modulational instabilities from the numerical solution of the dispersion relation, Eq.\eqref{disp_rel}. The growth rate map is populated by the normalized vector potential $a_0$ and the plasma density $n/n_c$. Since the growth rate of the backward Raman scattering branch (BRS) is not only higher than the modulational instability but also higher than the FRS branch of the Raman scattering, the parameter map essentially depicts the growth rate of the BRS instability. The region of maximum growth rate is close to the opaque region of plasma density. The growth rate first increases with $a_0$ and afterwards it decreases. The higher growth rate is due to higher plasma densities (without accounting for the relativistic transparency) while the decrease in the growth rate at high $a_0$ is attributed to the relativistic mass of electrons. This is expected since the lowering of the  growth rate due to relativistic electron mass effects can be countered by increasing the plasma density. Also the RR force does not strongly affect the growth rate of the BRS instability~\cite{Kumar:2013aa}. Thus, the growth rate of the BRS peaks for intermediate values of $a_0$ and decreases afterwards. Since, the enhancement due to the RR force in the growth rate of the BRS instability is smaller and it cannot counter the lowering caused by the relativistic mass effects. Due to this we do not show the result without the RR force, because the difference is not visible in the log$_{10}$ scale. Fig.\ref{fig:maxkmap} shows the maximum $k$ number of any unstable perturbation with (upper panel) and without (lower panel) the RR force. At low $a_0$ and $n_0$, there is no change in the wavenumber and it is depicting the maximum wavenumber of the BRS instability as expected. For high $a_0$ and $n_0$, a shift towards higher $k$ number is observed. The excitation of higher $k$ number suggests that the modulation instability can imprint fluctuations at smaller wavelengths on the laser pulse envelope during its propagation in a plasma.

\section{Spatial analysis of the dispersion relation}\label{spA}

\begin{figure}
\centering
\subfloat[\label{sfig:testa}]{%
  \includegraphics[width=0.38\textwidth,height=0.28\textwidth]{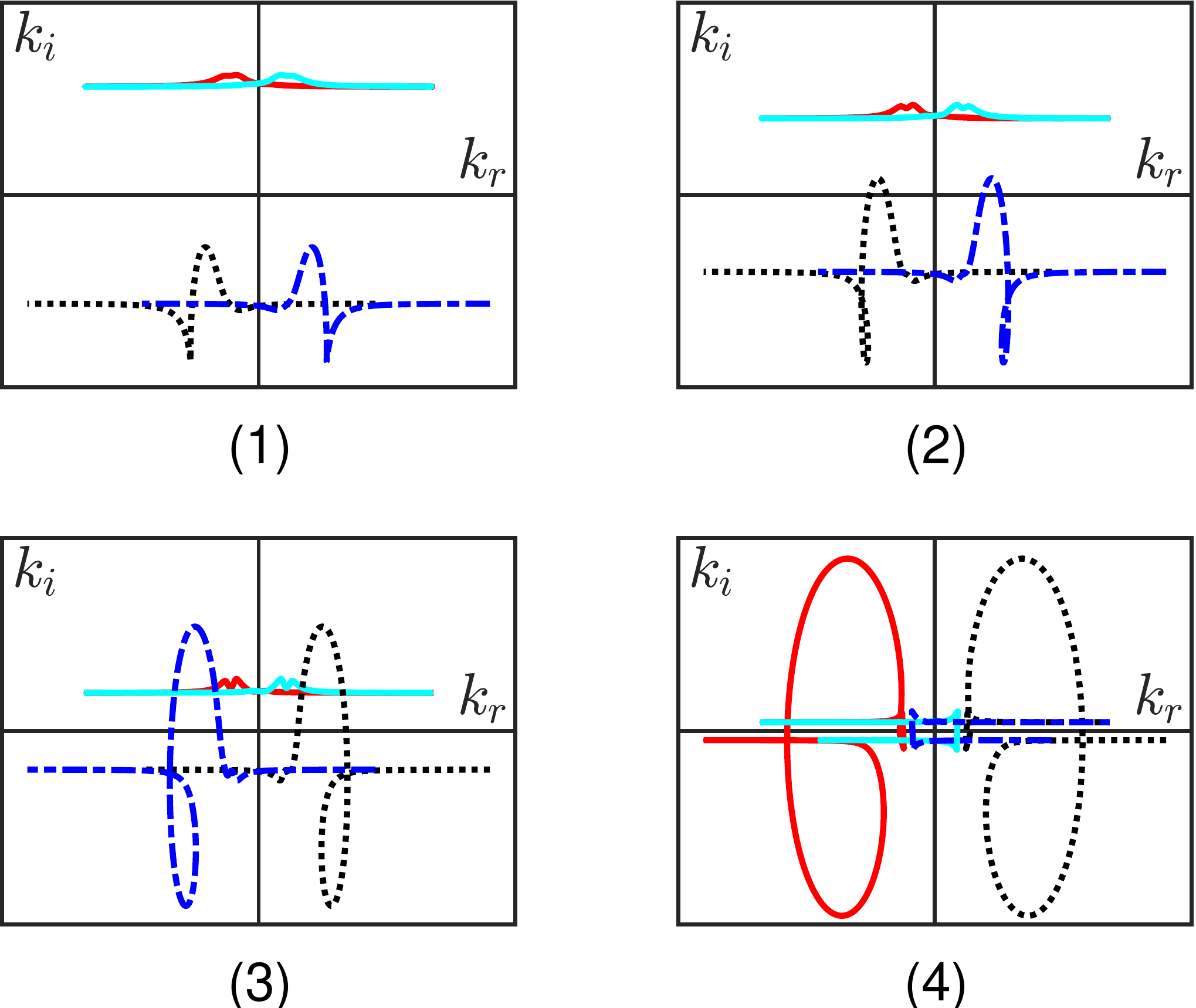}%
}\vfill
\centering
\subfloat[\label{sfig:testa}]{%
  \includegraphics[width=0.38\textwidth,height=0.28\textwidth]{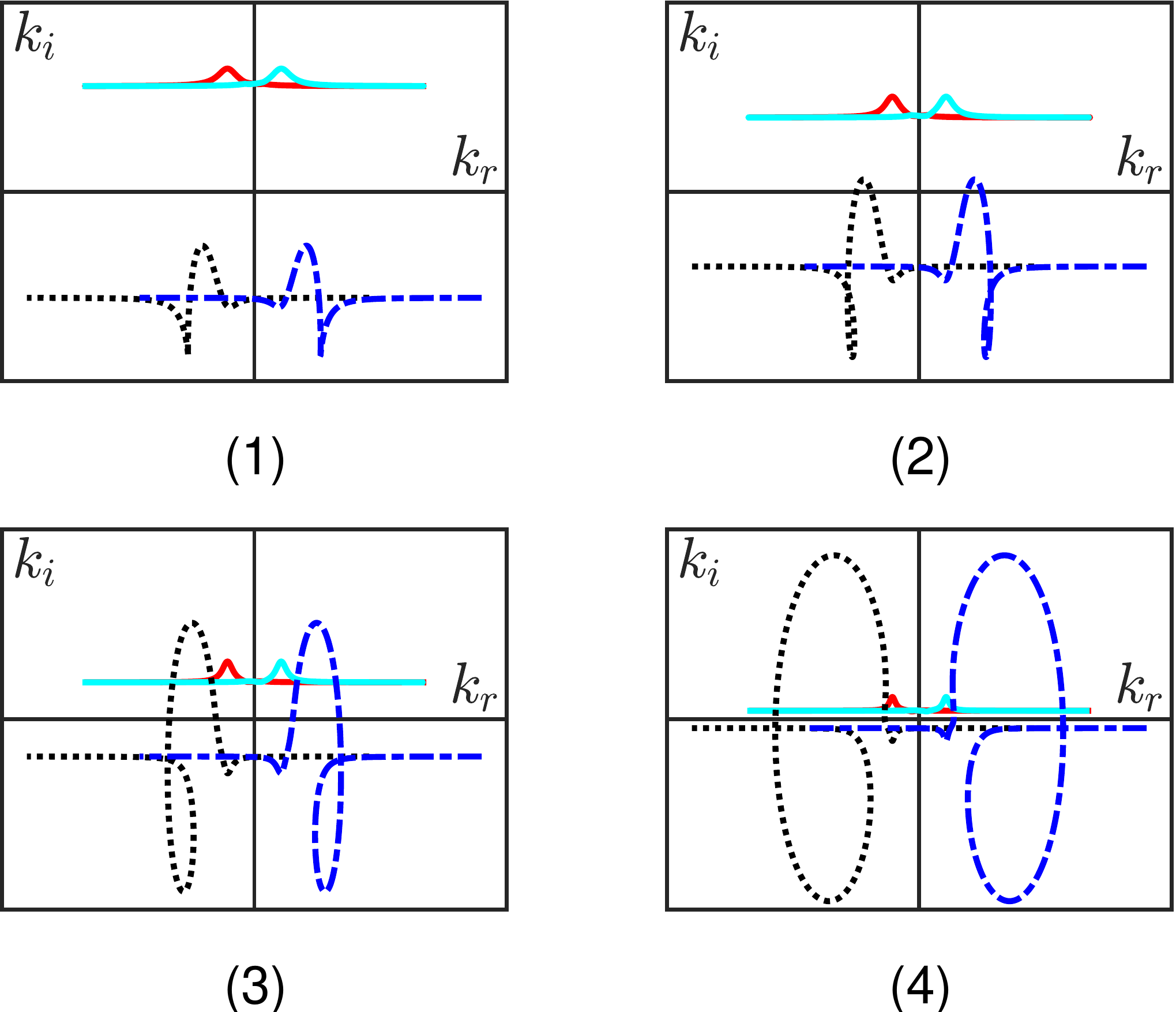}%
}
\caption{Solution of dispersion relation Eq.\eqref{disp_rel} in a complex $k$-plane with (a) and without (b) radiation reaction force. Panels $(1),(2),(3)$ and $(4)$ are for $\Im(\omega)=0.4,\,0.3\,,0.2$ and $0.1$ respectively in each case. The other parameters are $a_0=250, n=10\,n_c$. Both real and imaginary parts of the wavevector $k$ are normalized with $k_0$.}
\label{fig:Dk}
\end{figure}

\begin{figure}
\centering
\subfloat[\label{sfig:testa}]{%
  \includegraphics[width=0.38\textwidth,height=0.28\textwidth]{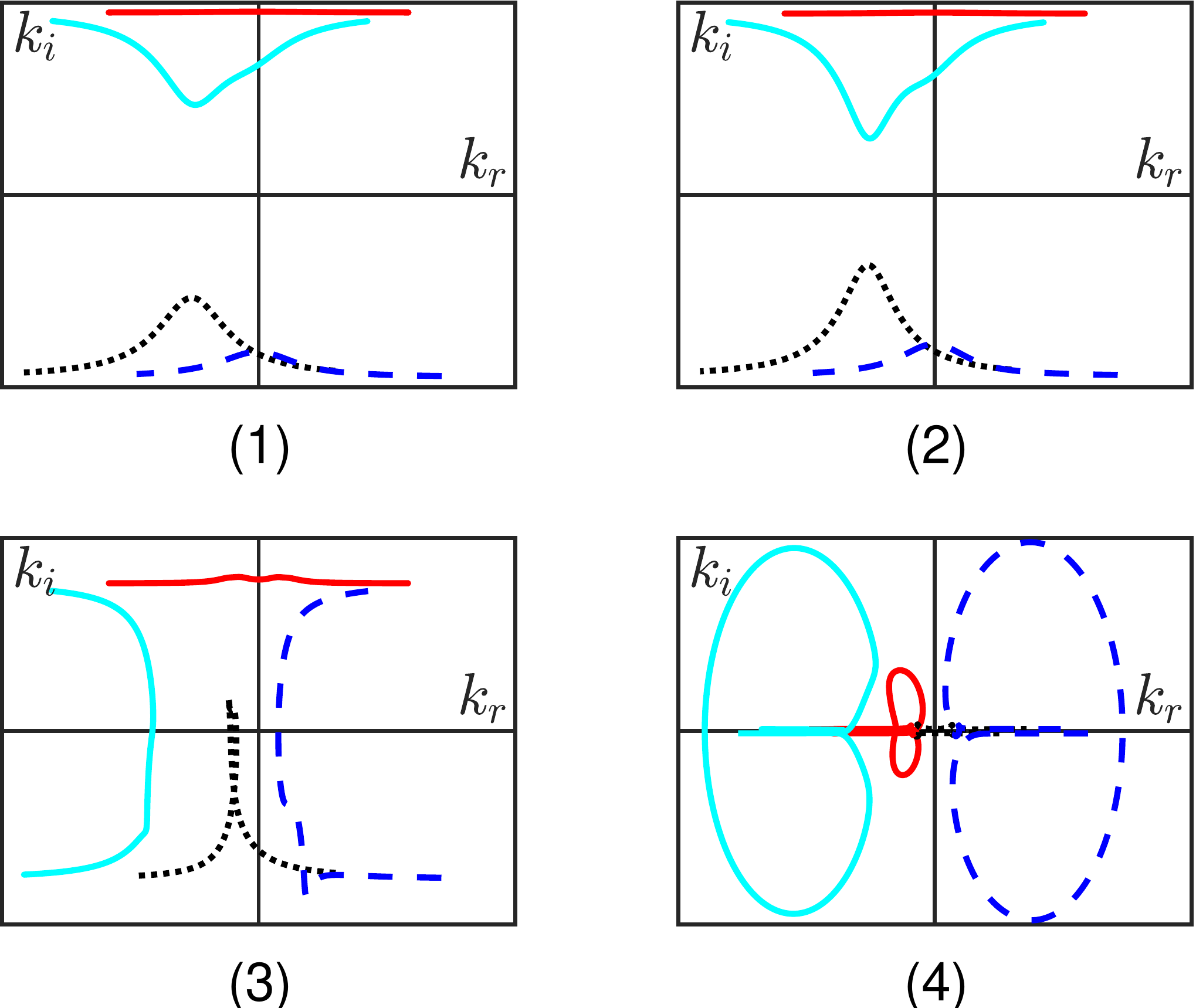}%
}\vfill
\centering
\subfloat[\label{sfig:testa}]{%
  \includegraphics[width=0.38\textwidth,height=0.28\textwidth]{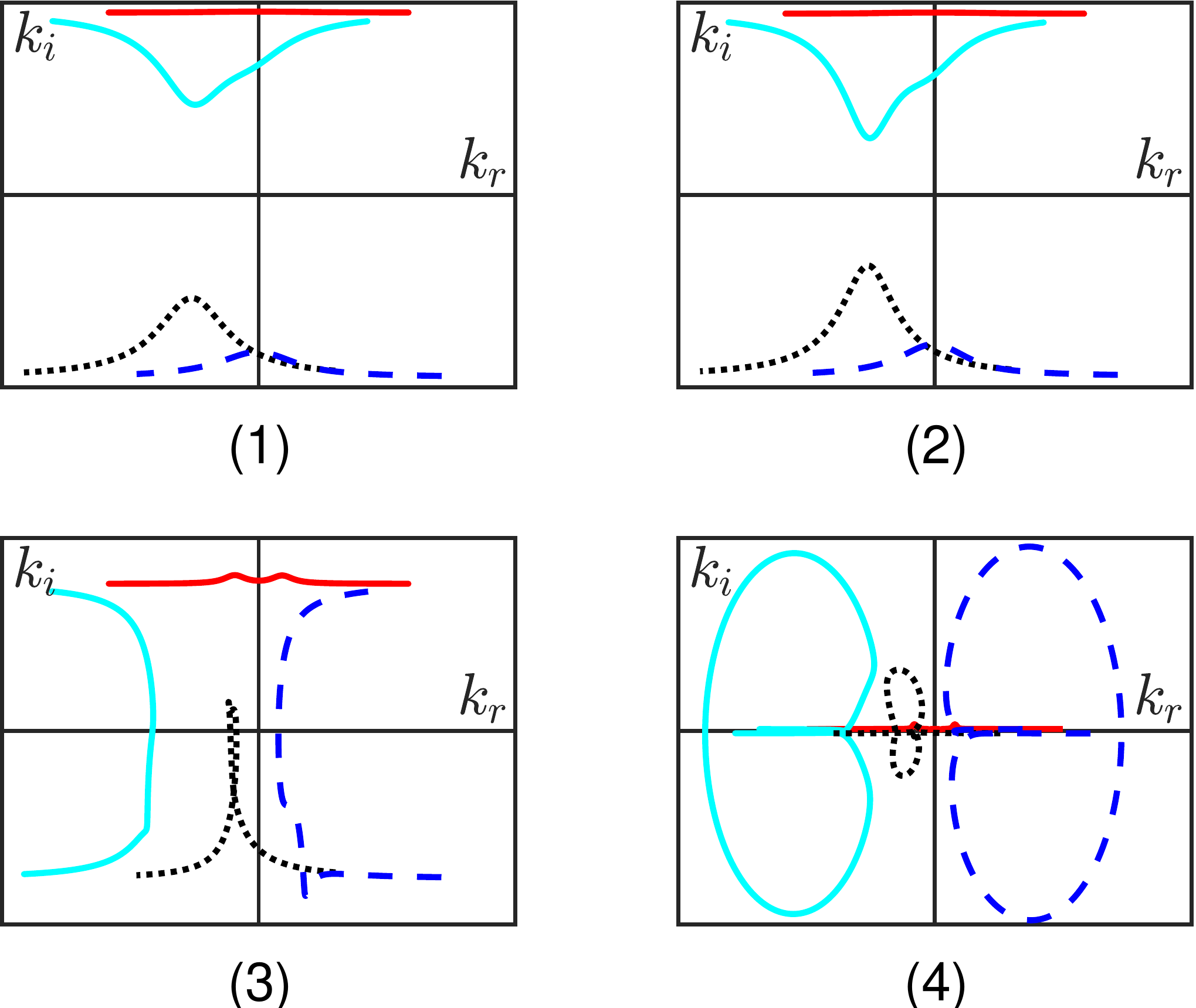}%
}
\caption{Solutions of equation, $D-\partial D/\partial k=0$, together with $\partial \omega/\partial k=0$ condition in a complex $k$-plane with (a) and without (b) radiation reaction force. Panels $(1),(2),(3)$ and $(4)$ are for $\Im(\omega)=2.0,\,1.8\,,0.5$ and $0.05$ respectively in each case. The other parameters are $a_0=250, n=10\,n_c$. Both real and imaginary parts of the wavevector $k$ are normalized with $k_0$.}
\label{fig:D-delD}
\end{figure}

Apart from carrying out the temporal analysis of the dispersion relation, we also carry out the spatial analysis to determine the evolution of an unstable perturbation in a plasma. Determining the nature of an unstable perturbation into absolutely and convectively instabilities \textcolor{black}{in a plasma} starts from the seminal works of Peter Sturrock~\cite{Sturrock:1994aa} and R. Briggs~\cite{Briggs:1964aa}. The procedure to determine the nature of instability essentially pertains to solving the dispersion relation \eqref{disp_rel} for a given $\omega$ and keeping track of the different $k$-branches in the complex $k$-plane. On varying the $\omega$, precisely $\Im(\omega)$, $k$-branches can move in the $k$-plane. Fig.\ref{fig:Dk} shows the evolution of branches in a complex $k$-plane with (a) and without (b) the RR force on varying the $\Im(\omega)$. We see branches in the lower half $k$-plane [branch with the \textcolor{black}{dot-dashed blue (dot-dashed dark grey) line]} move to the upper-half $k$-plane. This signifies the presence of a convective instability.  If for $\Im(\omega)>0$, there is also a collision between the branches from upper-half [$k_{+}(\omega)$] and the lower-half [$k_{-}(\omega)$] in the upper-half of the complex $k$-plane (as seen in the panel (4) of Fig.\ref{fig:Dk}), then we may have an absolute instability. The collision point is the saddle-point where the group velocity of the mode is zero \emph{i.e} $\partial \omega /\partial k =0$. If one of the branches briefly crosses the $k_r$-axis and upon colliding with a branch in the upper-half $k$-plane returns to the lower-half plane, we term it as the convective instability; as also seen in bottom-right sub-figure (4) of Fig.\ref{fig:Dk}(b).

\begin{figure}
\centering
\subfloat[\label{sfig:testa}]{%
  \includegraphics[width=0.32\textwidth,height=0.2\textwidth]{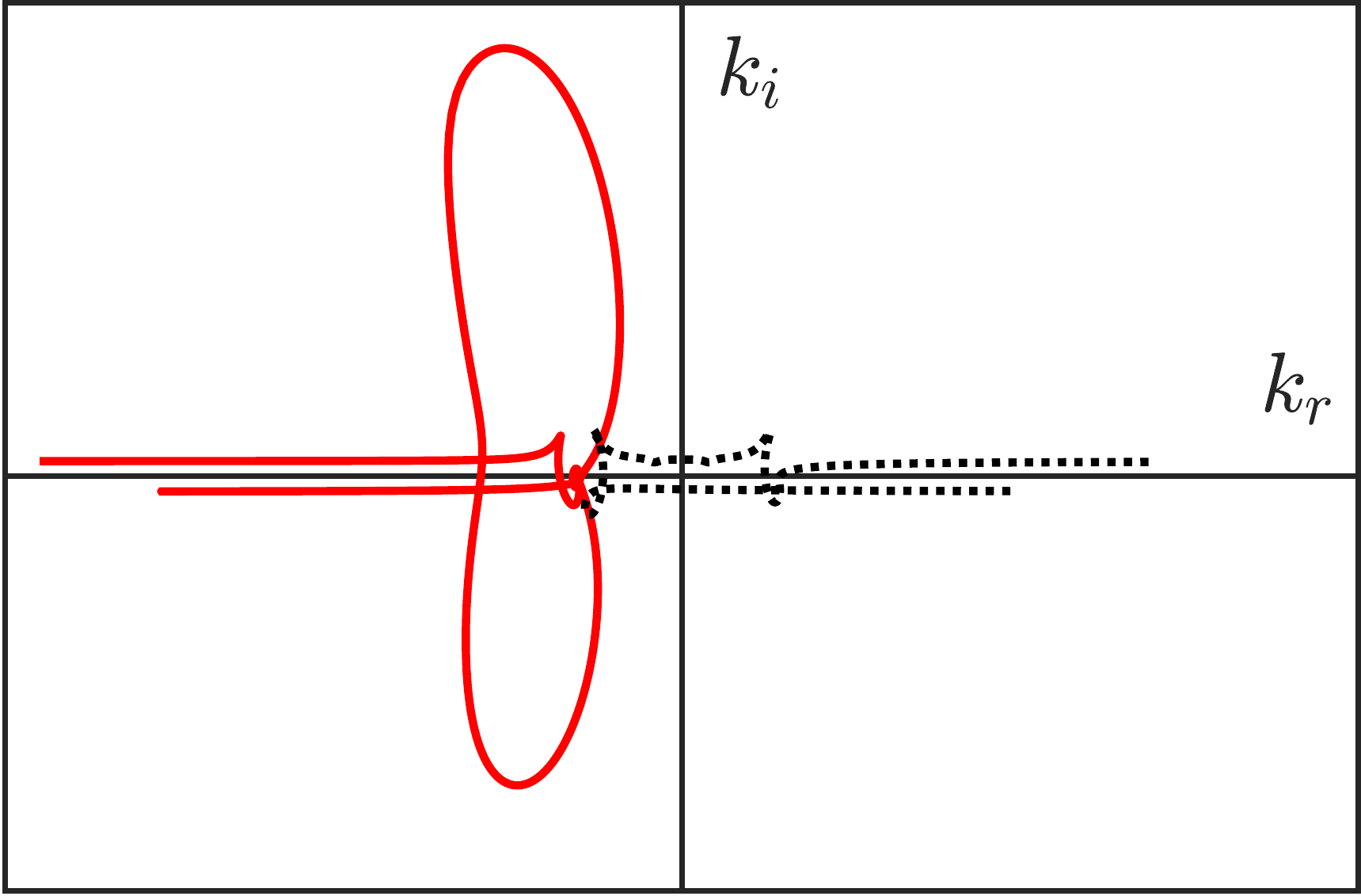}%
}\vfill
\centering
\subfloat[\label{sfig:testa}]{%
  \includegraphics[width=0.32\textwidth,height=0.2\textwidth]{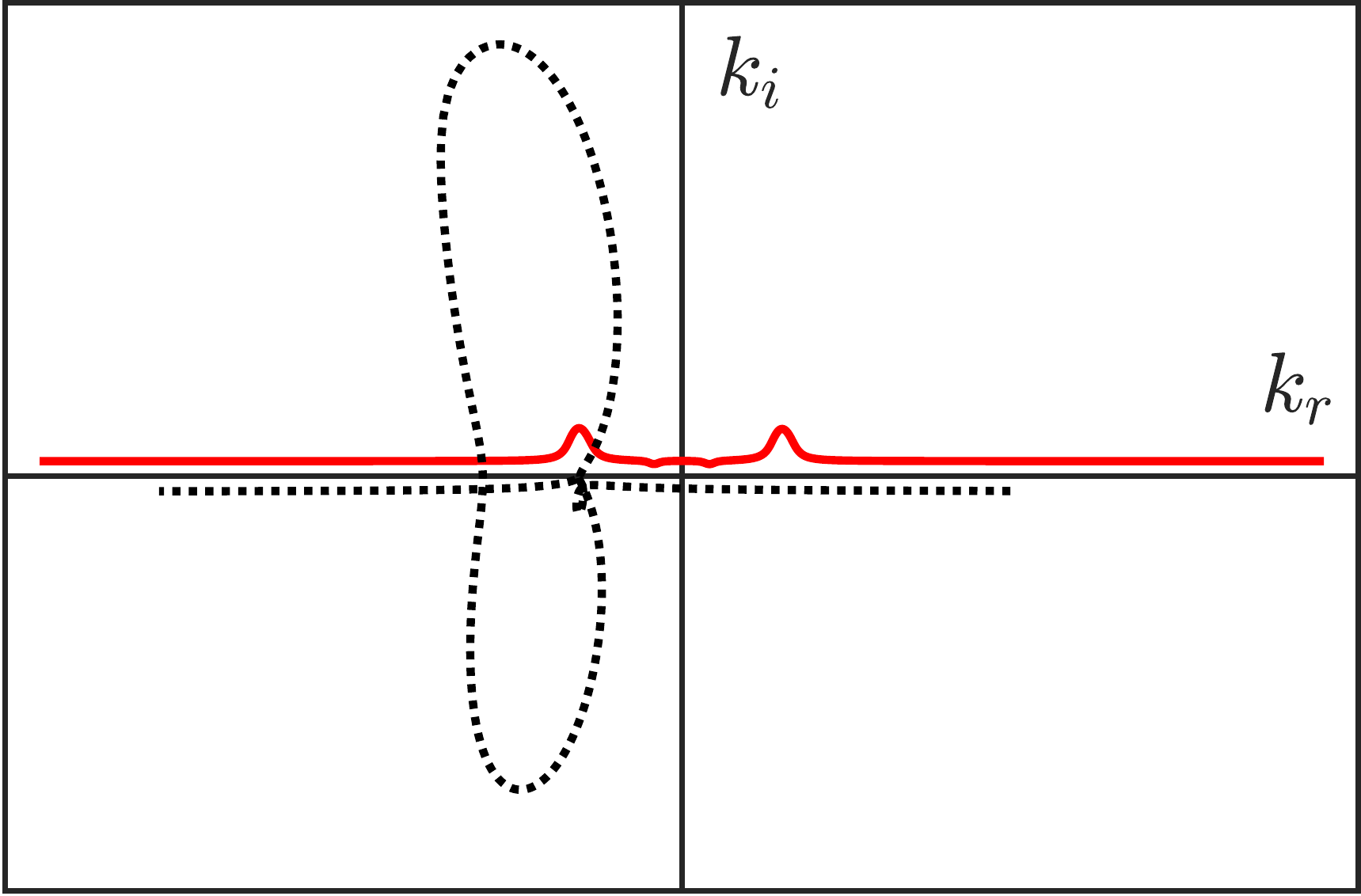}%
}
\caption{Magnified view of the panel (4) of Fig.\ref{fig:D-delD} with (a) and without radiation reaction force. It can be seen that one branch of the $k$-solution has crossed the $\Re(k)$-axis and is in the upper-half plane, signifying the absolute instability as seen in Fig.\ref{fig:Dk}. Other parameters are same as in Fig.\ref{fig:D-delD}. Both real and imaginary parts of the wavevector $k$ are normalized with $k_0$.}
\label{fig:twomode}
\end{figure}

The formal procedure to examine the nature of the instability, based on the causality and Green function formalism of a perturbation evolution, had been developed by  A. Bers~\cite{Bers:1985aa}, which has also been successfully employed to study the nature of the Raman and modulational instabilities of a laser pulse in an underdense plasma~\cite{couairon:3434,*Grismayer:2004aa}. In order to further check our results, we also use the same procedure as employed in Ref.~\cite{couairon:3434}. In order to find the saddle points, we solve two equations  $D(\omega_0,k_0)=0$ and $\partial D (\omega_0,k_0)/\partial k =0$  together with the zero group velocity condition $\partial \omega /\partial k =0$ at given $\Im(\omega), a_0$ and $n_0$ and then varying one parameter as stipulated in Ref.~\cite{couairon:3434}. We find a saddle-point if the two branches coming from different half $k$-planes cross the $k_r$-axis and collide in the upper half-plane for a finite $\Im(\omega)>0$. Fig.\ref{fig:D-delD} shows the $k$-branches for $a_0=250$ and $n=10 n_c$.  Analogous to the panels in Fig.\ref{fig:Dk}, we see that in the case of RR force, one of the $k$-branch (\textcolor{black}{dotted} black line) moves from the lower-half to the upper-half of the complex $k$-plane, colliding with the branch  [\textcolor{black}{solid red (dark grey) line]} from the upper-half $k$-plane, signifying the existence of an absolute instability. This movement can be clearly seen in Fig.\ref{fig:twomode}, which compares the two branches of the last two panels \textcolor{black}{(panel 4 in each case)} of Fig.\ref{fig:D-delD} with (a), and without (b) the RR force. 

\subsection{Parameter map for absolute and convective instabilities}\label{conV}

\begin{figure}[!t]
\centering
\subfloat[\label{sfig:testa}]{%
  \includegraphics[width=0.38\textwidth,height=0.25\textwidth]{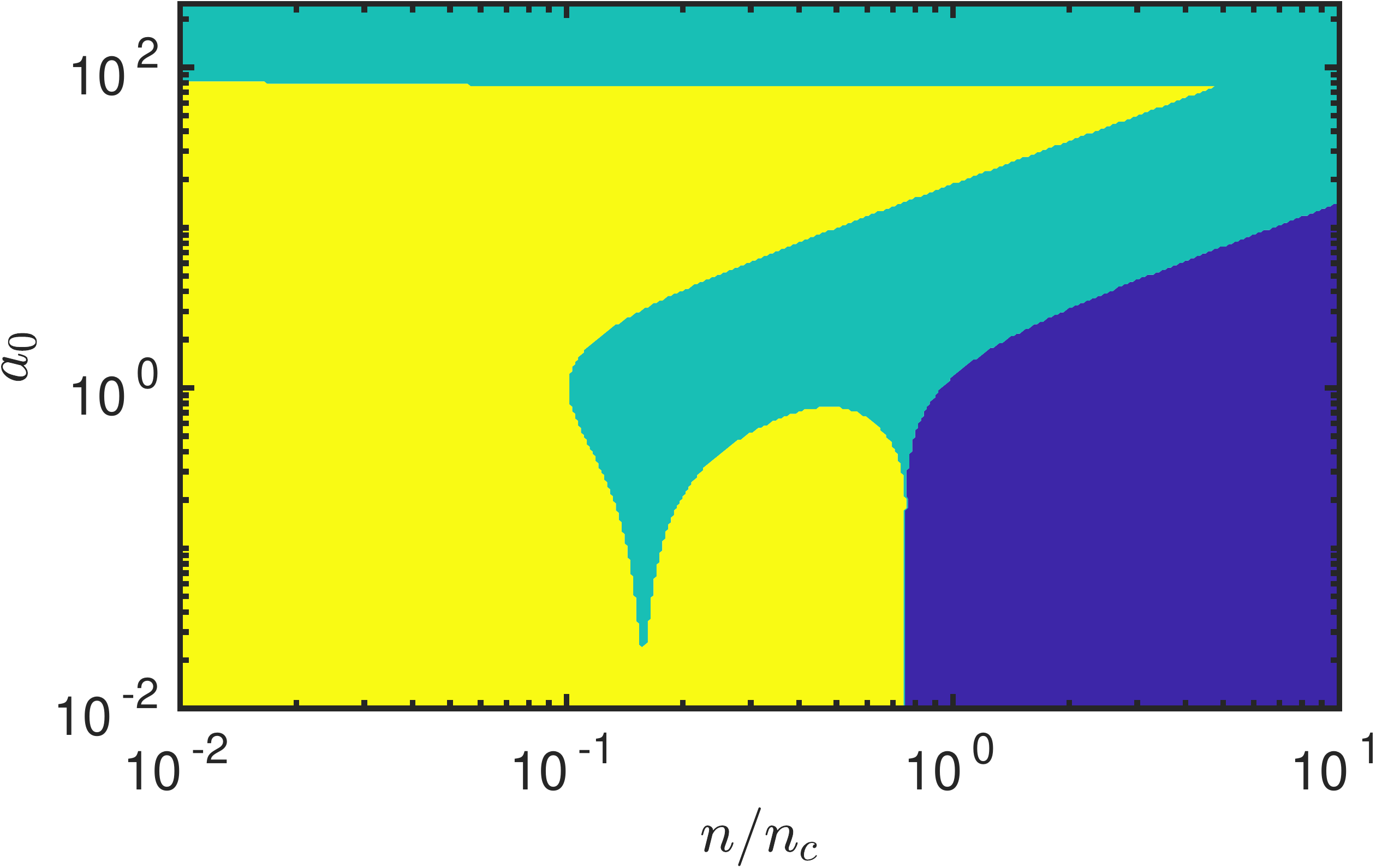}%
}\vfill
\subfloat[\label{sfig:testa}]{%
  \includegraphics[width=0.38\textwidth,height=0.25\textwidth]{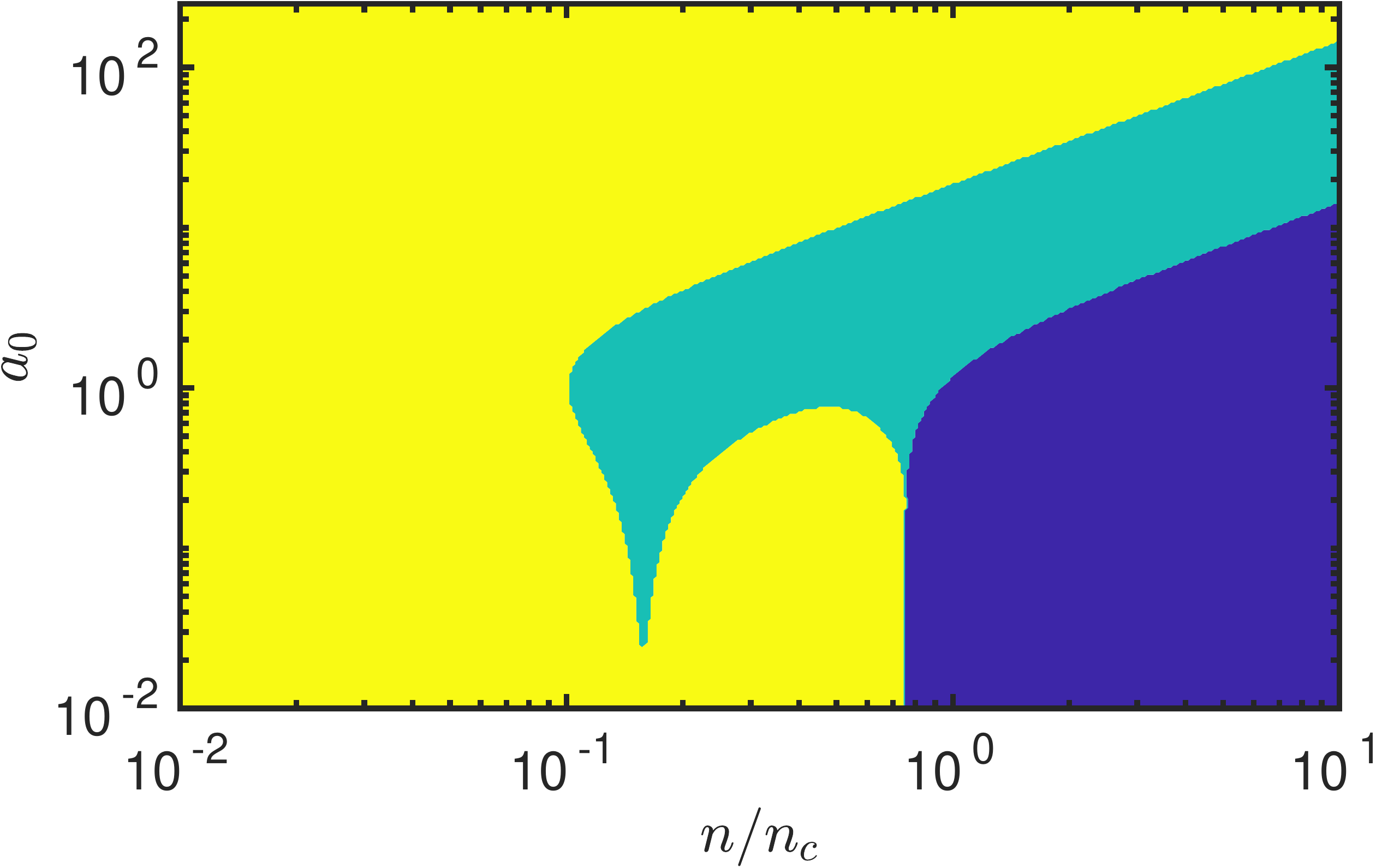}%
}
\caption{Parameter map showing the regions of absolute and convective instabilities with (a) and without (b) the inclusion of the radiation reaction force. Green \textcolor{black}{(light grey shading)} and yellow \textcolor{black}{(light shading)} colors represent the absolute and convective regions, respectively. Blue color \textcolor{black} {(dark grey shading)} denotes the relativistically opaque regions of plasma density.}
\label{fig:abs_conv_map}
\end{figure}

The movement of the branches in complex $k$-plane, \textcolor{black}{which determines the nature of the convective and absolute instabilities, depends on $a_0$ and $n_0$}. Hence, we plot a parameter map, spanned by the normalized vector potential $a_0$ and the density $n/n_c$, describing the boundaries of the absolute and convective instabilities in Fig.\ref{fig:abs_conv_map} with (upper panel) and without (lower panel) the RR force. For this, we solve the dispersion relation Eq.\eqref{disp_rel} for $\Im(\omega)=0.005\,\omega_0$ and $\Re(\omega)=[-7,7]$ and varying the pairs of ($a_0,n_0$). Our method is slightly different than the one adopted in Ref.~\cite{couairon:3434} but in the non-relativistic limit it reproduces the same parameter map as in Refs.~\cite{couairon:3434,*Grismayer:2004aa} as seen in Fig.\ref{fig:abs_conv_map}(b). It is clear that for higher $a_0$, where the radiation reaction force effects are important, the nature of the instability changes from convective to absolute.  We further cross-check our results with the procedure used in Ref.~\cite{couairon:3434} for  \textcolor{black}{different values} of $a_0$ and $n_0$ corresponding to the upper region of Fig.\ref{fig:abs_conv_map} (a) to ensure that we do have the existence of an absolute instability. We find the same behaviour as in Fig.\ref{fig:D-delD}(a) confirming the validity of our procedure and modification in the nature of a perturbation due to the RR force.  The presence of an absolute instability triggers a self-sustained oscillation which survives after the excitation of the instability, leading to the enhancement in the reflection coefficient of the laser pulse~\cite{couairon:3434}. Since the RR force changes the nature of the instabilities from convective to absolute, higher reflection and lower transmission of an ultra-relativistic laser pulse in a plasma are expected. This is counter-intuitive. Thus, the RR force appears to have significant impact not only on the growth rates of different modes of Raman and modulational instabilities but it also changes the nature of the plasma perturbation propagation. 

\section{conclusions and discussions}\label{conC}

We have numerically studied the effect of the radiation reaction force on the electronic parametric instabilities, particularly focusing on the branches of the Raman type instabilities. We have carried out both temporal and spatial analysis of the dispersion relation. We find that the RR force significantly changes the nature of the Raman branches of the parametric instabilities. Temporal analysis of the dispersion relation reveals that the RR force causes both the FRS and the BRS branches of the Raman instabilities to merge in a single branch while the spatial analysis of the dispersion relation suggests that the RR force changes the nature of instability from convective to absolute at high $a_0$. The merging of the FRS and the BRS modes of the Raman branch into a single mode implies that the wakefield acceleration in this regime is robust since the BRS instability has a higher growth rate and the onset of the BRS also implies the onset of the FRS due to the merging of these two modes in a single branch. Moreover, the trapping of the electrons due to the BRS can be helpful for the wakefield acceleration scheme in this regime. Since the RR force causes the nature of the perturbation to be absolute in the high $a_0$ regime, and the existence of an absolute instability implies the generation of self-sustained oscillations, this can lead to higher reflection of the laser pulse in this regime. It can have deleterious impact on the planned experiments to study the quantum effects in ultra-intense lasers since the higher reflection of the laser pulse in this regime can mitigate quantum effects to be observed experimentally~\cite{Mackenroth:2019aa,Marklund:2006aa,Raicher:2014aa}.

\textcolor{black}{In this paper, we have considered the laser pulse to be circularly polarized as it makes the analytical treatment of the formalism tractable. Qualitatively, the results are also applicable to the linearly polarized laser pulse case after appropriately adjusting the value of the normalized vector potential $a_0$. However, the linearly polarized light generates stronger harmonics in this ultra-relativistic regime~\cite{Sakharov:1997aa}. Generation of these harmonics can broaden the well-defined peaks of forward and backward Raman scatterings at $k \approx k_0$ and $k\approx 2k_0$ respectively. As also seen from Figs.\ref{sfig:testc} and \ref{fig:maxgrowth}, the largest growth rates of these two modes can be a fraction of laser frequency $\omega_0$. Thus the typical $e$-folding time for these modes to grow is on the electronic time scale. However, as the RR force causes the different modes of the Raman instability to merge in one branch [see Fig.\ref{sfig:testc}], the growth rates of this continuous Raman branch vary in $k$-space and can be really small for lower $k$-numbers as seen in Fig.\ref{sfig:testc}. Due to the relativistic mass variation of electrons, the electronic time scale also becomes larger \emph{e.g.} for $a_0=100$, the electronic time scale becomes larger by an order of magnitude. Thus, the ion motion can affect this Raman branch at $k$ numbers where the growth rate is smaller (lower $k$ numbers) and corresponding $e$-folding time is on the  ion motion time scale even though, the direct effect of the RR force on the ion motion is negligible.  It implies that the inclusion of the ion motion is expected to only significantly modifies the evolution of different modes with $k$ number for which the growth rate is really small (lower $k$ number). Modes with well-defined and larger $k$-numbers \emph{e.g} forward and backwards Raman scatterings are unaffected by the ion motion. Thus, the modulational branch of the instability can be significantly modified, at lower $k$ numbers, by the ion motion. Moreover, the ion motion can also excite new modes such as Brillouin scattering and coupling of these modes with electronic modes may be facilitated by the RR force. }

\textcolor{black}{Apart from the RR force, other factors such as collisional damping and Landau damping can also affect the electronic parametric instabilities of the laser pulse. In particular, collisions can reduce the growth rate of the electronic parametric instabilities but have shown to help distinguishing between the absolute and convective nature of the instabilities~\cite{guerin:2807,*Quesnel:1997aa,*Quesnel:1997ab,Esarey:2009fk}. We have not considered these factors in our calculations which is justified for the following reasons. Due to ultra-relativistic motion of plasma electrons, the collisions frequency, which scales as $\nu \propto \upsilon^{-3}$ becomes smaller. Here $\upsilon$ is the velocity of one of the two colliding plasma particles~\cite{Trubnikov:1965aa}. On the other hand, the RR force becomes stronger in ultra-relativistic regime and it also acts as effective collisions causing the phase shift between the electrons motion and the oscillations of the laser vector potential. Thus, the determination of the absolute vs convective nature of the instability becomes clearer with the inclusion of the RR force as seen in Figs.\ref{fig:Dk}, \ref{fig:D-delD} and \ref{fig:abs_conv_map}. Moreover, we consider a cold plasma and the phase velocity of the electromagnetic pulse in a cold plasma is superluminal. This precludes the Landau damping of the plasma wave. Vlasov simulations of ultra-intense laser pulse interaction with a plasma are suitable to verify the results presented in this paper. }

\acknowledgments{This work presented here encompasses the bachelor thesis of Ferdinand Gleixner submitted to the University of Heidelberg. \textcolor{black}{We thank Christoph Keitel for the appreciation of this work.} }


%

\end{document}